
\documentstyle{amsppt-n}
\magnification=1200
\hsize=15.5truecm
\vsize=22.5truecm
\font\citefont=cmbx9

\def\B{{\Cal B}}
\def\Cc{{\Cal C}}

\def\Rc{{\Cal R}}

\def\E{{\Cal E}}

\def\F{{\Cal F}}
\def\O{{\Cal O}}

\def\L{{\Cal L}}
\def\M{{\Cal M}}

\def\Zc{{\Cal Z}}
\def\Z{{\Bbb Z}}
\def\R{{\Bbb{R}}}
\def\C{{\Bbb{C}}}

\def\b#1,#2{B_L^{#1\vert#2}}

\def\pd#1,#2{\dfrac{\partial#1}{\partial#2}}
\def\sh#1#2{\hbox{$\Cal #1  #2 $}}

\def\Ext{\operatorname{Ext}}
\def\hom{\sh{H}om\,}
\def\rest #1,#2{{#1}_{\vert #2}}
\def\iso{\kern.35em{\raise3pt\hbox{$\sim$}\kern-1.1em\to}
         \kern.3em}
\def\longiso{\kern.7em{\raise3pt\hbox{$\sim$}\kern-1.5em
              \longrightarrow}\kern.3em}

\def\gcoor#1#2,#3#4{(#1^1,\allowmathbreak\dots,#1^{#2},\allowmathbreak
                    #3^1,\allowmathbreak\dots,#3^{#4})}

\def\proof{\noindent {\it Proof.} \ }
\def\dim{\operatorname{dim}}
\def\cite#1{[{\citefont #1}]}
\def\bigsearrow{\hbox{\bigsym \char '46}}

\font\bigsym=cmsy10 scaled\magstep3
\def\fd#1{R^{#1}\pi_\ast}
\def\intXY{\int_{X/Y}\,}
\def\Oast#1{\O_{#1}^\ast}
\def\fpr{\fd{1}\Oast{X}}
\def\pic{\operatorname{Pic}}
\def\picXY{\pic(X/Y)}

\def\cub{{\vrule width4pt height6pt depth1.5pt\hskip1pt}}
\noTagsNumbers
\nosubheadingnumbers
\TagsOnRight
\def\Nc{{\Cal N}}
\def\obligedskip{\vbox{\vskip4mm}}
\topmatter
\title
LINE BUNDLES OVER FAMILIES OF (SUPER) \\ RIEMANN SURFACES.
I: THE NON-GRADED CASE
\dag\\ \\ (March 1991 --- revised  September 1991 and October 1992)
\endtitle
\author
U\. Bruzzo\ddag\ {\rm and}\ J.A\. Dom\'\i nguez P\'erez\P
\endauthor
\affil
\ddag\thinspace Dipartimento di Matematica, Universit\`a
di Genova, Italia \\ \P\thinspace Departamento de
Matem\'atica Pura y Aplicada, \\ Universidad de Salamanca, Espa\~na
\endaffil
\address{(\ddag) Dipartimento di Matematica, Universit\`a
di Genova, Via L. B. Alberti 4, 16132 Genova, Italia. E-Mail: {\smc
bruzzo\@matgen.ge.cnr.it}}
\address{(\P) Departamento de Matem\'atica Pura y Aplicada,  Universidad
de Salamanca, Plaza de la Merced 1-4, 37008 Salamanca,
Spain}
\abstract{A first step towards a systematic theory of relative line
bundles
over SUSY-curves is presented. In this paper we deal with the case of
relative line bundles over families of ordinary Riemann surfaces.
Generalizations of the Gauss-Bonnet theorem and of the flatness theorem
for
line bundles are discussed.}
\thanks{\dag\ Research partly supported by the
joint  CNR-CSIC research project `Methods and applications of
differential
geometry in mathematical phys\-ics,' by `Gruppo Nazionale per la Fisica
Matematica' of CNR, by the Italian Ministry for University and Research
through
the research project `Metodi geometrici e probabilistici in fisica
matematica,'  and
by the Spanish CICYT through the research project `Geometr\'{\i}a de las
teor\'{\i}as gauge.'}   \endtopmatter  \document
\heading Introduction \endheading
Polyakov's approach to the computation of quantum scattering amplitudes
in the
bosonic string theory admits a nice geometric interpretation in terms of
integrals over the moduli space of Riemann surfaces
\cite{1}.
One expects that
superstrings (strings with bose and fermi degrees of freedom) can be
dealt with in
a similar way by introducing a suitable ${\Bbb Z}_2$-graded analogue  of
a
Riemann surface.

Physical arguments \cite{5} suggest that the
specification of a `super Riemann surface' should include:\par
\noindent 1. an ordinary Riemann surface $X$; \par
\noindent 2. a spin structure on $X$; \par
\noindent 3. a set of gravitino fields on $X$.

A convenient geometric setting for dealing with structures of this kind
is
provided by the Berezin-Le\u\i tes-Kostant graded manifolds
(or supermanifolds).
If one considers a (1,1) dimensional complex analytic graded manifold
$(X,{\Cal B}_X)$, then $(X,(\B_X)_0)$ is just an ordinary Riemann
surface;
$(\B_X)_1$ is in general a line bundle over $X$, and one can require
that it is
a spin structure on $X$. In order to accomodate the last datum,
i.e\. a set of  gravitino fields, one can in a sense consider {\sl a
family of
super Riemann surfaces}, also called {\sl SUSY-curve} \cite{12,13}.

The study of the geometry of these objects involves the extension  of
several constructions which are encountered in the usual theory of
Riemann
surfaces. We center our attention on some facts concerning (relative)
line
bundles over SUSY-curves; in particular, we deal with the extension of
two
results: the Gauss-Bonnet theorem, and the fact that a holomorphic line
bundle
on a Riemann surface is flat if and only if its Chern class vanishes.
This is
done in two steps: the first, which is dealt with in the present paper,
is the
generalization of these results to the case of a family of ordinary
Riemann
surfaces; the second deals with the graded case, and  forms the object
of a
further paper \cite{3}.

It should be noticed that some material we present here can be found in
some
classical works by Mumford and others; however, the techniques used
there are
quite different from those we employ here, which reduce to some complex
geometry and basic homological algebra. This makes the extension to the
graded setting quite straightforward.

Let us describe briefly the contents of this paper. After a cursory
presentation of the basic definitions, in Section 2 we develop the
notions of
relative de Rham theory and of fiberwise integration over a family of
smooth
manifolds; a relative Serre duality for families of complex analytic
manifolds
is also stated. In Section 3 we introduce, basically following
Grothendieck
\cite{7,8},  the notions of Picard sheaf and relative Picard group
of a family of complex manifolds; also the concept of Chern class finds
a
natural generalization to the relative context. A relative flatness
property is
stated, and it is proved that a section of the Picard sheaf is flat if
and only
if its relative Chern class vanishes. Moreover, we show that the
relative Chern
class can be represented in terms of the fiberwise integral
of a curvature form.

The main results of this paper and of the following one were announced
without
proofs in the report \cite{2}.
The graded version of some of them already appeared in the paper
\cite{6}; however, the transition from the absolute to the relative case
is not obtained --- as there claimed ---
simply by replacing the sheaf cohomology functor with
the higher direct image functor. Indeed, the category of families
of graded manifolds is essentially different
from that of single graded manifolds \cite{14,17}.
Also, the definition of relative flatness as given in
\cite{6} seems to be inadequate, and does not allow for a satisfatory
discussion of the validity of  the flatness  theorem for SUSY-curves.

\obligedskip\heading Families of smooth and complex analytic manifolds
\endheading \subheading{Preliminaries}
We shall denote by $(X,\O_X)$ a complex analytic manifold, and by
$(X,\Cc_X)$
a real differentiable manifold, where $\Cc_X$ is the sheaf of $\C$-
valued
smooth functions on $X$. All manifolds will be assumed to be connected.
Every $n$-dimensional complex manifold
determines an underlying $2n$-dimensional real manifold; if
$(z_1,\dots,z_n)$
are local coordinates in $(X,\O_X)$, then  $(z_1,\dots,z_n,\bar
z_1,\dots,\bar
z_n)$ are local coordinates for $(X,\Cc_X)$.
We shall use the notation $(X,\Rc_X)$ interchangeably for a real or
complex manifold, denoting in either case the local coordinate systems
by $\{z_i\}$. The sheaf $\sh Der\,X$ is the sheaf of derivations of
$\Rc_X$,
and $\Omega^1_X$ is the sheaf of differentials of $\Rc_X$, i.e\.
the $\Rc_X$-dual module of $\sh Der\,X$; these are locally free,
and any local coordinate system $\{z_i\}$ induces local bases
$\{\partial/\partial z_i\}$ of $\sh Der\,X$ and $\{dz_i\}$ of
$\Omega_X^1$.
The corresponding vector bundles are the tangent bundle $TX$ and the
cotangent bundle $T^\ast X$, respectively.
{}From the inverse function theorem one obtains that a set $\{z_i\}$
of sections of $\Rc_X$ is a local coordinate system around $x\in X$
if and only if the differentials of those sections at $x$, i.e\.
the elements $\{d_xz_i\}$, are a basis for $T^\ast_xX$.

A manifold morphism $\pi\colon (X,\Rc_X)\to (Y,\Rc_Y)$ is said to be
a {\sl submersion} if for all $x\in X$ the induced morphism
$\pi_x^\ast\colon T^\ast_{\pi(x)}Y\to T^\ast_xX$ is injective, or,
equivalently, if for all $x\in X$ there are open neighborhoods $U$ of
$x$ and
$V$ of $\pi(x)$, with $\pi(U)\subset V$,
such that (1) the
natural map $\pi^\ast\colon\Rc_Y(V)\to\Rc_X(U)$ is
injective; and (2) every coordinate system $\{w_i\}$
in $V$ can be completed to a coordinate system $\{w_i,z_j\}$ in $U$.

The morphism $\pi\colon (X,\Rc_X)\to (Y,\Rc_Y)$ is said to
be {\sl proper} if it is such in the sense of topological
spaces;  {\sl flat} if $\pi_\ast \Rc_X$ is a flat sheaf of
$\Rc_Y$-modules. In the complex analytic case,
if $\pi_\ast\O_X$ is coherent (which for instance happens
when $\pi$ is proper via Grauert's cohomology base change theorem)
the flatness
of $\pi$ implies that $\pi_\ast\O_X$ is locally free over $\O_Y$.

For all $y\in Y$, the fiber $X_y=\pi^{-1}(y)$ of a proper morphism $\pi$
is a compact
space, which is endowed with a manifold structure given by the
structure sheaf
$$\Rc_{X_y}=\rest{(\Rc_X/\widehat{\frak m}_y)},{X_y}\simeq
\rest{\Rc_X},{X_y}\otimes_{(\Rc_Y)_y}k(y)$$
where $\hat{\frak m}_y$ is the ideal of $\Rc_X$
generated by the inverse image of ${\frak m}_y$; here ${\frak m}_y$
denotes
the maximal ideal of $(\Rc_Y)_y$, and also the naturally associated
sheaf
of ideals of $\Rc_Y$. Moreover, $k(y)$ is the
residual field at $y$, i.e\. $k(y)=(\Rc_Y)_y /{\frak m}_y\simeq\C$.
The manifold $(X_y,\Rc_{X_y})$ can also be regarded
as the fibered product $X\times_Y\{y\}$.

We say that a morphism $\pi\colon (X,\Rc_X)\to (Y,\Rc_Y)$
has {\sl universally connected fibers\/} if
$\pi_\ast\Rc_X\simeq\Rc_Y$; this implies indeed that  all fibers $X_y$
are
connected, and that this property is preserved under any change of the
basis
$(Y,\Rc_Y)$.

A {\sl family of manifolds} is a proper
submersion $\pi\colon (X,\Rc_X)\to (Y,\Rc_Y)$. In the complex
analytic case we also assume that $\pi$ is flat and
has universally connected fibers.
The {\sl relative dimension} of the family is the number
$\dim X -\dim Y$, which corresponds to the dimension of each fiber.
A complex analytic family of relative dimension 1 is called
a {\sl family of Riemann surfaces}.
A {\sl relative coordinate system} is a local coordinate system
$\{w_i,z_j\}$ on $X$ such that $\{w_i\}$ is a local coordinate system
on $Y$.

If $\pi:(X,\O_X)\to (Y,\O_Y)$ is a family of complex manifolds
of relative dimension $n$ there is a vanishing theorem
$\fd{k}\O_X=0$ for all $k>n$, where $\fd{k}$ denotes
the $k$th higher direct image functor associated with $\pi_\ast$.

A {\sl morphism of families\/} between $\pi'\colon (X',\Rc_{X'})\to
(Y',\Rc_{Y'})$
and $\pi\colon (X,\Rc_X)\to (Y,\Rc_Y)$ --- which we shall briefly
denote by $f\colon(X'/Y')\to(X/Y)$ ---
is a pair of morphisms $f\colon (X',\Rc_{X'})\to(X,\Rc_{X})$
and $f'\colon (Y',\Rc_{Y'})\to (Y,\Rc_{Y})$
such that the following diagram commutes:
$$
\CD
(X',\Rc_{X'}) @>f>>  (X,\Rc_{X}) \\
@V\pi' VV @VV\pi V \\
(Y',\Rc_{Y'}) @>>f'>  (Y,\Rc_{Y}) \endCD \quad.
$$

The {\sl sheaf of relative derivations} of a family
$\pi\colon (X,\Rc_X)\to (Y,\Rc_Y)$,
denoted $\sh Der\,(X/Y)$, is the subsheaf of $\sh Der\,X$ whose
elements  vanish on $\Rc_Y$, that is, the following exact sequence
holds:
$$ 0 @>>> \sh Der\,(X/Y) @>>>    \sh Der\,X @>>>
\pi^\ast \sh Der\,Y @>>> 0\,.$$
The dual module of $\sh Der\,(X/Y)$, denoted $\Omega^1_{X/Y}$, is
the {\sl sheaf of relative differentials} of the family, and one
has an exact sequence
$$ 0 @>>> \pi^\ast\Omega^1_Y @>>> \Omega_X^1 @>>> \Omega^1_{X/Y} @>>>
0\,.$$
Any relative coordinate system $\{w_i,z_j\}$ induces local bases
$\{\partial/\partial z_j\}$ of
$\sh Der\,(X/Y)$ and $\{dz_j\}$ of $\Omega_{X/Y}^1$.

For every fiber $X_y=\pi^{-1}(y)$ there are identifications
$$\Omega^1_{X_y}\simeq\rest{(\Omega^1_{X/Y}/\widehat{\frak m}_y
\Omega^1_{X/Y})},{X_y}\simeq
\rest{\Omega^1_{X/Y}},{X_y}\otimes_{(\Rc_Y)_y}k(y)\,,$$
namely, the differentials of the fibers can be obtained from the
relative differentials by restricting to the fibers and taking values;
in this sense, if $\{w_i,z_j\}$ is a relative coordinate system on the
family,
the $\{dz_j\}$'s are a local basis for $\Omega^1_{X_y}$, and $\{z_j\}$
is a local coordinate system for the fiber.

A family $\pi\colon(X,\Cc_X)\to(Y,\Cc_Y)$ of real manifolds
is said to be {\sl orientable} if the relative cotangent bundle   ---
namely,
the vector bundle associated with the locally free module
$\Omega^1_{X/Y}$
--- is orientable, which amounts to saying that the highest exterior
power of $\Omega^1_{X/Y}$ has a global nowhere vanishing section.
Intuitively, this means that all fibers are orientable in a compatible
way.
A family of real manifolds which underlies a family of complex
manifolds is orientable, and carries a canonical orientation,
exactly in the same way as the real manifold underlying a complex
manifold is canonically oriented.

\subheading{Relative de Rham theory}
Let $\pi\colon(X,\Cc_X)\to(Y,\Cc_Y)$ be a family of real manifolds
of relative dimension $m$; we denote by $\Omega^k_{X/Y}$ the
$k$th exterior power of $\Omega^1_{X/Y}$ over $\Cc_X$,
and call {\sl relative $k$-forms} its sections. The exterior
differential $d\colon\Cc_X\to\Omega^1_X$ induces, by composition with
the projection $p\colon \Omega^1_X\to\Omega^1_{X/Y}$, a {\sl relative
differential} whose kernel is $\pi^{-1}\Cc_Y$ by definition. By
extending in the usual way to the exterior algebra, one obtains
a relative differential $d_r\colon \Omega^{k-1}_{X/Y}\to\Omega^k_{X/Y}$
for all $k>0$, which makes the following diagram commute:
$$\CD
\Omega^{k-1}_X @>d>> \Omega^k_X \\
@V p VV @VV p V \\
\Omega^{k-1}_{X/Y} @>>d_r > \Omega^k_{X/Y} \endCD\quad.
\tag 2.1$$
On the other hand, the relative differential induces in each fiber
a differential operator which coincides with the ordinary exterior
differential.

We denote by  $\Zc^k_{X/Y}$ the sheaf of closed relative $k$-forms,
i.e\. the kernel of the morphism
$d_r\colon \Omega^k_{X/Y}\to\Omega^{k+1}_{X/Y}$;
in particular, $\Zc^m_{X/Y}\equiv\Omega^m_{X/Y}$.
A `relative Poincar\'e lemma' holds, that is, the
following sequence of sheaves on $X$ is exact \cite{4,18}:
$$ 0 @>>> \pi^{-1} \Cc_Y @>>> \Cc_X @>{d_r}>> \Omega^1_{X/Y} @>{d_r}>>
\dots
@>{d_r}>> \Omega^m_{X/Y} @>>> 0 \,. \tag 2.2$$
So one has an acyclic  resolution of the sheaf $\pi^{-1}\Cc_Y$.
\proclaim{Definition}
The relative de Rham sheaf  of degree $k$ is the sheaf over $Y$
$$DR^k_{X/Y}\equiv \frac{\pi_\ast \Zc^k_{X/Y}}{{d_r \pi_\ast\Omega^{k-
1}_{X/Y}}}
\,.$$
\endproclaim

Equivalently, $DR^k_{X/Y}$ is the sheaf associated with the presheaf
$$ V   \rightsquigarrow    \frac{\Zc^k_{X/Y} (\pi^{-1}(V))}
{d_r \Omega^{k-1}_{X/Y} (\pi^{-1}(V)) }\,.\tag 2.3$$
Thus, $DR^k_{X/Y}$ is the $k$th cohomology sheaf of the complex
of sheaves over $Y$
$$0 @>>> \pi_\ast\pi^{-1} \Cc_Y @>>> \pi_\ast\Cc_X @>{d_r}>>
\pi_\ast\Omega^1_{X/Y} @>{d_r}>> \dots
@>{d_r}>> \pi_\ast\Omega^m_{X/Y} \,. $$
If $Y$ reduces to a point, $Y=\{y\}$, then $\Gamma(Y,DR^k_{X/Y})$
is the ordinary $k$th de Rham cohomology group of $X$.
\proclaim{Proposition} For each $k\geq 0$ there is a canonical sheaf
isomorphism
$$DR^k_{X/Y}\longiso \fd{k}\pi^{-1}\Cc_Y\,.$$
\endproclaim
\proof Given a resolution $\F^\bullet$ of the sheaf
$\pi^{-1} \Cc_Y$, there is a canonical morphism ${\Cal H}^k(\pi_\ast
\F^\bullet)\to\fd{k}\pi^{-1} \Cc_Y$ (abstract de Rham theorem);
applying this to the resolution (2.2), one obtains a sheaf morphism
$DR^k_{X/Y}\to \fd{k}\pi^{-1}\Cc_Y$. In order to prove that this is
bijective, we can restrict to the stalks, thus getting
$$(DR^k_{X/Y})_y\simeq\frac{ \Gamma(X_y,\Zc^k_{X/Y})}{
d_r\,\Gamma(X_y,\Omega^{k-1}_{X/Y})},\qquad
(\fd{k}\pi^{-1}\Cc_Y)_y\simeq H^k(X_y,\pi^{-1}\Cc_Y)\,.$$
The  corresponding morphism between the right-hand sides is bijective
by the de Rham theorem for sheaf cohomology, in that the
sequence (2.2), when restricted to $X_y$, yields an acyclic resolution.
\qed\enddemo
Quite obviously, if $Y=\{y\}$, Proposition 2.2 reduces to the
ordinary de Rham theorem for $X$.

We wish now to investigate the relationship occuring between
the relative de Rham cohomology of a family and the ordinary
de Rham cohomology of the total space $X$.

\proclaim{Lemma}
There is a commutative diagram of $\C$-modules
$$\CD
\Gamma(X,\Zc^k_X) @>>> H^k(X,\C) @>>> \Gamma(Y,\fd{k}\C)\\
@V p VV @VV p V @VVV \\
\Gamma(Y,\pi_\ast\Zc^k_{X/Y}) @>>> \Gamma(Y,DR^k_{X/Y})
@>\sim>> \Gamma(Y,\fd{k}\pi^{-1}\Cc_Y)
\endCD \quad.
$$
\endproclaim
\proof Let us consider the square on the left;
in view of the definition of relative differential, for
any open $V\subset Y$ there is a commutative diagram
$$\CD
\Omega_X^{k-1}(\pi^{-1}(V)) @> d >> \Zc_X^{k}(\pi^{-1}(V))
@>>> \frac{\dsize\Zc_X^{k}(\pi^{-1}(V))}{\dsize
d\,\Omega_X^{k-1}(\pi^{-1}(V))} \\
@V p VV @VV p V @VV p V \\
\Omega_{X/Y}^{k-1}(\pi^{-1}(V)) @> d_r >> \Zc_{X/Y}^{k}(\pi^{-1}(V))
@>>> \frac{\dsize\Zc_{X/Y}^{k}(\pi^{-1}(V))}{\dsize
d_r\,\Omega_{X/Y}^{k-1}(\pi^{-1}(V))}
\endCD$$
which yields a commutative diagram of presheaves. One concludes by
passing
to the associated sheaves in the bottom line and taking global sections.

The square on the right is induced by the diagram
$$\CD
\pi_\ast\Zc^k_X/d\,\pi_\ast\Omega^{k-1}_X @>>> \fd{k}\C \\
@V p VV @VVV \\
DR^k_{X/Y} @>>> \fd{k}\pi^{-1}\Cc_Y
\endCD \quad;
\tag 2.4$$
here $\pi_\ast\Zc^k_X/d\,\pi_\ast\Omega^{k-1}_X $ is the quotient
{\sl pre\/}sheaf, and the morphism in the first line is the
canonical map to the associated sheaf.
So, it is enough to prove that the
diagram (2.4) is commutative, and this in turn can be proved
by restricting to the stalks.  One then considers the
commutative diagram
$$
\CD
0 @>>> \C @>>> \Omega^\bullet_X \\
@. @VVV @VV p V \\
0 @>>> \pi^{-1}\Cc_Y @>>> \Omega^\bullet_{X/Y} \endCD\quad,
$$
applies the direct image functor $\pi_\ast$, and takes cohomology,
thus obtaining diagram (2.4) with the presheaf in the upper
left corner replaced by the corresponding sheaf; the result
we are looking for follows by composing with the canonical morphism
mapping
the presheaf to the sheaf.
\qed\enddemo
\subheading{Fiberwise integration}
Relative forms of top degree on a oriented family of real manifolds
$\pi\colon(X,\Cc_X)\to(Y,\Cc_Y)$ can be integrated `along the
fibers' to yield a function on the base space $Y$.
More precisely, if the family has relative dimension $m$,
fiberwise integration is a sheaf morphism
$$ \intXY\colon\pi_\ast\Omega^m_{X/Y}\to\Cc_Y\,,$$
defined as follows: if $\omega$ is a section of
$\pi_\ast\Omega^m_{X/Y}$, for all $y$ in the domain of $\omega$
one denotes by $\bar\omega_y\in\Gamma(X_y,\Omega^m_{X_y})$
the image of the germ $\omega_y \in\Gamma(X_y,\Omega^m_{X/Y})$; then
$$\bigl(\intXY\omega\bigr)(y)=\int_{X_y}\,\bar\omega_y\,.$$
The integral in the right-hand side is the usual integral over
a compact orientable manifold, and
the fact that the left-hand side depends differentiably on $y$
can be checked in local coordinates. If $Y=\{y\}$, fiberwise integration
reduces to ordinary integration over $X$.
\proclaim{Proposition} {\rm (Stokes theorem)}
For any section $\tau$ of $\pi_\ast\Omega^{m-1}_{X/Y}$ one has
$$\intXY  d_r\,\tau=0\,.$$
Thus, fiberwise integration induces a morphism
$$\intXY\colon DR^m_{X/Y}\to \Cc_Y\,.\tag 2.5 $$
\endproclaim
\proof In view of the definitions of fiberwise integration and
relative differential, the first assertion reduces to Stokes theorem
in each fiber. Then one has a morphism from the presheaf (2.3) into
$\Cc_Y$,
and one concludes by factorizing through the associated sheaf.
\qed\enddemo
\proclaim{Proposition} The morphism (2.5) is bijective.
\endproclaim
\proof The demonstration is the same as in the absolute case \cite{16},
for the functions on $Y$ are constant as far as fiberwise integration
is concerned.
\footnote{This is expressed by the equality $\intXY (\pi^\ast
f)\omega = f \intXY\omega$.}
Indeed, the family being orientable, the morphism $\intXY$
is not zero, i.e\. there is a relative $m$-form $\omega$
such that $\intXY\omega\neq 0$. One has to prove that the class
$[\omega]$ is a generator of $DR^m_{X/Y}$, in the sense that
for any other section $\omega'$ of $\pi_\ast\Omega^m_{X/Y}$
there are sections  $f$ of $\Cc_Y$ and $\tau$ of $\pi_\ast\Omega^{m-
1}_{X/Y}$
such that $\omega'=f\omega+d_r\,\tau$. This fact is first proved
locally on a coordinate cover (so that it reduces to the case
$Y=\R^s$, $X=\R^s\times C$, where $C$ is a compact orientable manifold)
and then globalized by means of a partition of unity argument.
\qed\enddemo
Fiberwise integration can be used to introduce a {\sl relative
Poincar\'e duality}. One first defines a $\Cc_Y$-bilinear sheaf
morphism
$$\fd{k}\pi^{-1}\Cc_Y  \otimes_{\Cc_Y} \fd{k'}\pi^{-1}\Cc_Y
\to \fd{k+k'}\pi^{-1}\Cc_Y \tag 2.6$$
in the following way: $\fd{k}\pi^{-1}\Cc_Y$ is the sheaf
associated with the presheaf $V\rightsquigarrow H^k(\pi^{-1}(V),\pi^{-1}
\Cc_Y)$; a morphism like (2.6) is defined at the level of
presheaves as the cup product, and then extended to the associated
sheaves (by regarding the sheaves $\fd{k}\pi^{-1}\Cc_Y$ as
$DR^k_{X/Y}$, this corresponds to the wedge product of
relative forms).
If $k'=m-k$, by composing the morphism (2.6) with fiberwise
integration, one obtains a pairing
$$\fd{k}\pi^{-1}\Cc_Y  \otimes_{\Cc_Y} \fd{m-k}\pi^{-1}\Cc_Y
\to \Cc_Y\,, \tag 2.7$$
which is non-degenerate, as one can show for instance by restricting to
the stalks. In that case, the pairing (2.7) reduces to
$$H^k(X_y,\pi^{-1}\Cc_Y) \otimes_{(\Cc_Y)_y} H^{m-k}(X_y,\pi^{-1}\Cc_Y)
\to (\Cc_Y)_y\,.\tag 2.8$$
Since $\rest{\pi^{-1}\Cc_Y},{X_y}$ is the constant sheaf $(\Cc_Y)_y$,
from the universal coefficient theorem one has
$H^k(X_y,\pi^{-1}\Cc_Y)\simeq H^k(X_y,\C)\otimes_\C(\Cc_Y)_y$,
so that (2.8) is the pairing
$$H^k(X_y,\C) \otimes_\C H^{m-k}(X_y,\C) \otimes_\C (\Cc_Y)_y
\to (\Cc_Y)_y\,,$$
which is the $(\Cc_Y)_y$-bilinear extension of the Poincar\'e
duality on the fiber $X_y$, and is therefore non-degenerate.
\qed\enddemo

\subheading{Relative duality for analytic families}
Given a complex analytic family $\pi\colon$\break
$(X,\O_X)\to (Y,\O_Y)$,
one can introduce a {\sl relative Serre duality},
which we would like to recall without providing any proof
(cf\. \cite{9,10,15}).
Assume that the family has relative dimension $n$,
denote by $\kappa_{X/Y}$ the sheaf of relative holomorphic
$n$-forms, and let $\M$ and $\Nc$ be coherent
$\O_X$- and $\O_Y$-modules, respectively; then, denoting by $R$ the
operation
of taking the derived functor, there is a canonical
isomorphism  of $\O_Y$-modules
$$R\hom_{\O_Y}(R\pi_\ast\M
,\Nc)\simeq R\pi_\ast R\hom_{\O_X}(\M,\kappa_{X/Y}(-n)
\otimes_{\O_Y}\Nc)\,.$$
One has then a convergent spectral sequence
$$\E^{p,q}_2=\Ext^p_{\O_Y}(R^{n-q}\pi_\ast\M, \Nc)
\Longrightarrow \E^{p+q}_\infty =\Ext^{p+q}_{\pi}(\M,
\kappa_{X/Y}\otimes_{\O_Y}\Nc) \,, $$
so that there is a natural morphism\footnote{This morphism
is clearly bijective for any $\M$ whenever $\Nc$
is a coherent injective sheaf, or for any $\Nc$ if all higher direct
images $\fd{k}\M$ are locally free.}
$$\E^{0,q}_2=\hom_{\O_Y}(R^{n-q} \pi_\ast\M, \Nc)\to\E^q
_\infty =\Ext^q_{\pi}(\M,\kappa_{X/Y}\otimes_{\O_Y}\Nc)
\,.\tag 2.9$$
For $q=0$, one deduces an isomorphism
$\hom_{\O_Y}(R^n\pi_\ast\M,\allowmathbreak
\Nc)\simeq\allowmathbreak\pi_\ast\hom_{\O_X}(\M,
\allowmathbreak\kappa_{X/Y}
\otimes_{\O_Y}\Nc)$ for arbitrary $\M$ and $\Nc$, and
one obtains in particular
 $$\eqalign{R^n\pi_\ast\kappa_{X/Y}&\simeq
(\pi_\ast\O_X)^\vee\cr &  \simeq(\O_Y)^\vee \simeq \O_Y\,,\cr}\tag
2.10$$
where $^\vee$ denotes the dual module. The second isomorphism
depends on $\pi$ having universally
connected fibers, while the first does not.

\obligedskip\heading Relative theory of line bundles \endheading
\subheading{Picard sheaf and relative Picard group}
The notion of relative Picard group, in the sense we are going
to employ it, has been introduced by Grothendieck, both
in the algebraic \cite{7} and  analytic case \cite{8}.
We would like to discuss
that concept in the case of a family of complex manifolds,
by making explicitly all the assumptions valid in that framework.
Our aim is to classify `families of line bundles' over the fibers
of  the morphism $\pi\colon (X,\O_X)\to (Y,\O_Y)$, where such `families'
are, speaking heuristically, collections of line bundles, one on each
fiber
of $\pi$, with a certain compatibility condition. In order to do that,
it is quite natural, on the analogy of what happens in the absolute
case, to consider a `relative cohomology' of the sheaf $\O_X^\ast$
of invertible sections of $\O_X$. This relative cohomology is of course
provided by the higher direct image functor $\fd{\bullet}$.
\proclaim{Definition}
The Picard sheaf of the family $\pi\colon (X,\O_X)\to (Y,\O_Y)$ is the
sheaf
$\fd{1}\Oast{X}$. \endproclaim
The group $\Gamma (Y,\fpr)$ is the {\sl (restricted) relative Picard
group}
$\picXY$ (cf\. Grothendieck \cite{8});
if $Y=\{y\}$, the group $\picXY$ reduces to the ordinary Picard group
$\pic(X)$ of $X$. In the general case, the elements in the relative
Picard group can be, according to the intuitive description above
reported,
interpreted as follows (this brief discussion is taken from \cite{8}).
The sheaf $\fpr$ is associated with the presheaf
$V\rightsquigarrow \allowmathbreak H^1(\pi^{-1}(V),\Oast{X})\simeq
\allowmathbreak \pic(\pi^{-1}(V))$ (group of isomorphism classes of
line bundles over $\pi^{-1}(V)$). The specification of an element of
$\picXY$
is equivalent to the assignation of an open cover $\{V_i\}$ of
$Y$ and a collection  of line bundles $\{\L_i\}$, each
defined over $\pi^{-1}(V_i)$, such that  for all pairs
$i,\,j$, the bundles $\rest{\L_i},{f^{-1}(V_i\cap V_j)}$ and
$\rest{\L_j},{f^{-1}(V_i\cap V_j)}$ are locally isomorphic relative
to $V_i\cap V_j$, in the sense that any $y\in V_i\cap V_j$
has an open neighborhood $W\subset V_i\cap V_j$ such that
$\rest{\L_i},{f^{-1}(W)}\simeq\rest{\L_j},{f^{-1}(W)}$.

The fact that $\O_Y\simeq\pi_\ast\O_X$, together with the Leray
spectral sequence for the morphism $\pi$, gives rise to an exact
sequence
$$ 0 @>>> \pic (Y) @>>> \pic (X) @> \phi >> \picXY \,.$$
One denotes by $[\L]$ the image  of an
isomorphism class $\L \in \pic (X)$ under $\phi$.
The morphism $\phi$ is in general not surjective, unless $\pi$ admits
a global section. Moreover, $\phi$ is functorial, in the following
sense.
\proclaim{Proposition}
Given a morphism of complex analytic families $f\colon(X'/Y')\to(X/Y)$,
there is a commutative diagram
$$\CD
\pic (X) @> \phi >> \picXY\\
@Vf^\ast VV @VV f^\ast V \\
\pic (X') @>> \phi'> \pic (X'/Y')
\endCD\quad.\tag 3.1$$
\endproclaim
\proof It follows from the functoriality of the Leray sequence
\cite{11}.
\qed\enddemo
The morphism $f^\ast\colon\pic (X)\to \pic (X')$ is actually induced by
the inverse image of sheaves of $\O_X$-modules.
In the particular case where the morphism is
the immersion of a fiber, $i\colon(X_y/\{y\})\hookrightarrow (X/Y)$, the
diagram (3.1) and the identification $\pic (X_y/\{y\}) \simeq \pic
(X_y)$
entails $[\L]_y \simeq \L_y$, where $[\L]_y$ is the stalk at $y$ of the
section
$[\L] \in \pic (X/Y)$, and $\L_y \in \pic (X_y)$ is the line bundle
$\L_y = \allowmathbreak \rest{\L},{X_y}\otimes_{(\O_Y)_y}k(y)
\allowmathbreak\simeq
\allowmathbreak\rest{\L},{X_y}\otimes_{\rest{\O_X},{X_y}} \O_{X_y}$.

\proclaim{Definition} The relative Chern class $c_1(\lambda)$
of a section $\lambda$ of the relative Picard sheaf $\fd{1}\Oast{X}$
of the family $\pi\colon (X,\O_X)\to (Y,\O_Y)$ is minus its image
via the sheaf morphism
$$\fd{1}\Oast{X}\to\fd{2}\Z\tag 3.2$$
induced by the exact sequence
$$ 0 @>>> \Z @>>> \O_X @> {\exp 2\pi i} >> \O_X^\ast @>>>0 \,.\tag3.3$$
\endproclaim
{}From (3.2) one obtains a morphism
$c_1\colon \Gamma (Y,\fpr)\to \Gamma (Y,\fd{2}\Z )$, which we call
{\sl Chern class} as well; in the case  $Y=\{y\}$, this morphism reduces
to the ordinary Chern class $c_1\colon \pic(X)\to H^2(X,\Z )$.

\proclaim{Proposition} There is a commutative diagram
$$\CD
\pic (X) @>\phi >> \picXY \\
@V{c_1} VV @VV{c_1}V  \\
H^2(X,\Z) @>>>H^0(Y, R^{2}{\pi}_\ast\Z)
\endCD\quad.\tag 3.4$$
\endproclaim
\proof It follows from the definition of Chern class.
\qed\enddemo

\proclaim{Proposition}
Given a morphism of complex analytic families $f\colon(X'/Y')\to(X/Y)$,
there is a commutative diagram
$$\CD
\picXY @> f^\ast >> \pic (X'/Y') \\
@V{c_1} VV @VV{c_1}V  \\
\Gamma (Y,\fd{2}\Z) @>>f^\ast> \Gamma (Y', R^{2}{\pi'}_\ast\Z)
\endCD\quad.\tag 3.5$$
\endproclaim
\proof One considers the commutative diagram
$$ \CD
0  @>>> \Z @>>> f^{-1} \O_X     @> {\exp 2\pi i} >>
                               f^{-1}\O_X^\ast @>>> 0 \\
@.     @|     @VVV       @VVV            \\
0  @>>> \Z  @>>>  \O_{X'}  @> {\exp 2\pi i} >> \O_{X'}^\ast @>>> 0 \\
\endCD\quad;$$
by applying the higher direct image functor and taking global sections
one concludes.
\qed\enddemo
{}From (3.5) applied  in the case of the immersion of a fibre
$i\colon(X_y/\{y\})\hookrightarrow (X/Y)$,
one has an identification $(c_1([\L]))_y \simeq c_1(\L_y)$
for each element $[\L] \in \picXY$.

If $\pi\colon (X,\O_X)\to (Y,\O_Y)$ is a family of Riemann surfaces,
the Chern class $c_1\colon \allowmathbreak
\fpr \allowmathbreak\to \fd{2}\Z $ can be
considered as a section in $\fd{2}\C$, due to the following result.

\proclaim{Proposition} For a family of Riemann surfaces, the natural
injection $\Z\hookrightarrow\C$ induces an immersion
$$\fd{2}\Z \hookrightarrow \fd{2}\C \,,$$
so that $\Gamma (Y,\fd{2}\Z )\hookrightarrow \Gamma (Y,\fd{2}\C )\,.$
\endproclaim\proof It is enough to prove that for any $y\in Y$ the
morphism induced
on the stalks $(\fd{2}\Z)_y \to (\fd{2}\C)_y$  --- that is, the morphism
$H^2(X_y,\Z)\to H^2(X_y,\C)$ --- is injective.
This follows from the commutative diagram
$$
\CD
H^2(X_y,\Z) @>>>  H^2(X_y,\C) \\
@VVV @VVV \\
\Z @>\text{inj}>> \C \endCD
\quad,\tag 3.6$$
where the vertical arrows are the Poincar\'e duality isomorphisms, which
can be realized in $H^2(X_y,\C)$ in terms of the integral over $X_y$.
\qed\enddemo

\subheading{Relative flatness}
We wish to generalize to the relative situation the well-known fact that
a holomorphic line bundle on a Riemann surface is flat if and only if
its Chern class vanishes. We recall that a line bundle over
an $n$-dimensional analytic manifold $(X,\O_X)$
is said to be {\sl flat} if its transition functions
over a suitable trivializing open cover can be chosen so as to be
locally constant.
Equivalently, on can consider the immersion
$\C^\ast\hookrightarrow\O^\ast_X$, and then (the isomorphism classes of)
flat bundles are the elements in the image of the
induced morphism $H^1(X,\C^\ast )\to H^1(X,\O^\ast_X )$.
In the case of a family of analytic manifolds $\pi\colon (X,\O_X)\to
(Y,\O_Y)$, the `relative constants' (in the sense that their relative
differential vanishes) are the sections  of $\pi^{-1}\O_Y$, which
suggests the following definition.

\proclaim{Definition} A  section of
the relative Picard sheaf $\fd{1}\Oast{X}$ of the family
$\pi\colon (X,\O_X)\to (Y,\O_Y)$ is said to be flat if it lies in the
image
of the morphism
$$\fd{1}\pi^{-1}\O^\ast_Y\to\fd{1}\Oast{X}\tag 3.7$$
induced by $\pi^{-1}\O^\ast_Y\hookrightarrow\O^\ast_X\,.$
\endproclaim
\proclaim{Theorem} Let $\pi\colon (X,\O_X)\to (Y,\O_Y)$ be a
family of Riemann surfaces. Any $y\in Y$ has an open
neighborhood $V$ such that a section $\lambda \in \Gamma (V,\fpr)$ has
vanishing  relative Chern class if and only if it is  flat.
\endproclaim
\proof If $\kappa_{X/Y}$ denotes the sheaf of relative holomorphic
$1$-forms, on has a commutative diagram

$$ 
\CD
@.      @.    0     \\
@.     @.    @AAA            \\
@.      @.       \kappa_{X/Y}   \\
@.     @.    @AAA            \\
0  @>>> \Z @>>>  \O_X  @>>> \O_X^\ast @>>> 0 \\
@.     @|     @AAA       @AAA            \\
0  @>>> \Z @>>>  \pi^{-1}\O_Y  @>>> \pi^{-1}\O_Y^\ast @>>> 0 \\
\endCD\quad.$$
Applying the higher direct image functor, one obtains
$$ 
\CD
R^2\pi_\ast\O_X \\
@AAA \\
R^2\pi_\ast\pi^{-1}\O_Y \\
@A \alpha AA \\
R^1\pi_\ast\kappa_{X/Y} \\
@AAA \\
R^1\pi_\ast \O_X        @>>> R^1\pi_\ast \O_X^\ast        @> \dsize
{c_1} >>
               R^2\pi_\ast\Z    @>>>      R^2\pi_\ast\O_X   \\
@AAA @AAA @| @AAA\\
R^1\pi_\ast\pi^{-1}\O_Y @>>> R^1\pi_\ast\pi^{-1}\O_Y^\ast @>>>
               R^2\pi_\ast\Z    @>\beta>> R^2\pi_\ast\pi^{-1}\O_Y  \\
\endCD\quad.\tag 3.8$$
The relative dimension of the family $\pi\colon (X,\O_X)\to (Y,\O_Y)$
being $1$, one has\break $R^2\pi_\ast\O_X = 0$. If we prove that
$\alpha$ is bijective and $\beta$ is injective, diagram (3.8) yields
$$ 
\CD
0 \\
@AAA \\
R^1\pi_\ast \O_X  @>>> R^1\pi_\ast \O_X^\ast @> \dsize {c_1} >>
               R^2\pi_\ast\Z    @>>> 0   \\
@AAA @AAA \\
R^1\pi_\ast\pi^{-1}\O_Y @>>> R^1\pi_\ast\pi^{-1}\O_Y^\ast @>>> 0 \\
\endCD\quad;\tag 3.9$$
the two lines of this diagram are exact, which
implies our claim. Thus, the theorem reduces to the following two
Lemmas.
\qed\enddemo

\proclaim{Lemma} The epimorphism
$\alpha\colon R^1\pi_\ast\kappa_{X/Y} \to R^2\pi_\ast\pi^{-1}\O_Y $
is actually an isomorphism.
\endproclaim
\proof From Eq\. (2.10) one has  $R^1\pi_\ast\kappa_{X/Y}
\simeq  \O_Y$. Besides,
$(R^2\pi_\ast\pi^{-1}\O_Y)_y$ $\simeq H^2(X_y,\pi^{-1}\O_Y)
\simeq H^2(X_y,(\O_Y)_y)$, since $\rest{\pi^{-1}\O_Y},{X_y}$
is the constant sheaf $(\O_Y)_y$, and from the universal coefficient
theorem one has $(R^2\pi_\ast\pi^{-1}\O_Y)_y \simeq
H^2(X_y,\C)\otimes_\C(\O_Y)_y \simeq (\O_Y)_y $.
Taking the quotient by the maximal ideal one gets an
isomorphism $\alpha\colon\C\to\C$,
and by Nakayama's Lemma one concludes.
\qed\enddemo
One should remark that this results depends critically on
$\pi$ having universally connected fibers.
\proclaim{Lemma} The morphism
$\beta\colon R^1\pi_\ast\Z \to R^2\pi_\ast\pi^{-1}\O_Y $
is injective.
\endproclaim
\proof It is enough to prove that for any $y\in Y$ the morphism induced
on the stalk, $\beta\colon (R^2\pi_\ast\Z)_y \to
(R^2\pi_\ast\pi^{-1}\O_Y)_y $, is injective.
A computation similar to that of the previous as Lemma enables to write
this morphism as
$$\beta\colon H^2(X_y,\Z) \to H^2(X_y,\C)\otimes_\C(\O_Y)_y \,,$$
which can be identified with the cohomology morphism induced by
$\Z\hookrightarrow (\O_Y)_y\simeq \C\otimes_\C(\O_Y)_y$, $m\mapsto
m\otimes 1$.
Then the claim follows from the diagram (3.6).
\qed\enddemo

\subheading{Gauss-Bonnet theorem for families}
Let $(X,\O_X)$ be a complex analytic manifold, and let $\L$ be a line
bundle
on it. It is a classical result that  $c_1(\L)= \frac{i}{2\pi} [K]$,
where
$[K]$ is the de Rham cohomology class of a
(smooth) curvature form $K$ on $\L$. If $(X,\O_X)$ is a compact
Riemann surface, the isomorphism $\int_X\,\colon H^2(X,\C)\iso \C$
allows one
to express this result in the form $c_1(\L)= \frac{i}{2\pi} \int_X\, K$
(Gauss-Bonnet theorem).
If $\pi\colon (X,\O_X)\to (Y,\O_Y)$ is a family of complex manifolds,
and $\pi\colon (X,\Cc_X)\to (Y,\Cc_Y)$ is the underlying real family,
the projection $p\colon \Omega^2_X\to\Omega^2_{X/Y}$ between
the sheaves of smooth $2$-forms induces
a morphism $p\colon H^2(X,\C)\to \Gamma (Y,DR^2_{X/Y})$, and
one has an identification $p(c_1(\L))= \frac{i}{2\pi} p([K])$. By
applying
Lemma 2.3 and Propositions 3.4 and 3.6, one deduces
$$c_1[\L]= \frac{i}{2\pi} [p(K)] \in \Gamma (Y,DR^2_{X/Y})\,,\tag 3.10$$
where $c_1[\L] \in \Gamma (Y,\fd{2}\C)$ is regarded as a section in
$\Gamma (Y, \fd 2 \pi^{-1} \Cc_Y)$ $\simeq$\break
$\Gamma (Y,DR^2_{X/Y})$.
The relative $2$-form $p(K) \in \Gamma (Y,\pi_\ast \Zc_{X/Y}^2)
= \Gamma (X,\Zc_{X/Y}^2)$ can be interpreted as a `relative curvature'
in the following way.
\proclaim {Definition} Given a family of complex manifolds
$\pi\colon (X,\O_X)\to (Y,\O_Y)$ and a line bundle $\L$ over $(X,\O_X)$,
let $\Omega^1_{X/Y}$ be the sheaf of relative differentials of the
underlying real family; a (smooth) relative connection over
$\L$ is a morphism of $\C$-modules
$$\nabla_r\colon \L\to \Omega^1_{X/Y}\otimes_{\O_X}\L\,\tag 3.11$$
which satisfies the Leibniz rule $\nabla_r (f\sigma)=
d_rf\otimes\sigma + f\nabla_r (\sigma)$.
\endproclaim
The projection of ordinary differentials onto relative differentials
allows one
to regard any relative connection as induced by an ordinary connection.
\proclaim{Lemma} For any relative connection $\nabla_r$ on $\L$
there is an ordinary connection $\nabla$ such that the following
diagram commutes:
$$\CD
\L @>\nabla>> \Omega^1_X \otimes_{\O_X}\L \\
@. @V\nabla_r\bigsearrow\phantom{xxxxxxxxx} V p\otimes\operatorname{Id} V
\\
@. \Omega^1_{X/Y}\otimes_{\O_X}\L \\
\endCD\quad.\tag 3.12$$
\endproclaim
\def\nzero{\nabla^0}
\proof Fix a connection $\nzero$ on $\L$, and let $\nzero_r$
be the corresponding relative connection, $\nzero_r=(p\otimes
\operatorname{Id})\nzero$. For any relative connection
$\nabla_r$ let $\alpha$ be a counterimage of $\nabla_r-\nzero_r$
under $p\otimes
\operatorname{Id}$. Then $\nabla=\nabla^0+\alpha$ is the
desired connection.\qed\endproclaim
We extend $\nabla_r$ to a morphism $\Omega^1_{X/Y}\otimes_{\O_X}\L\to
\Omega^2_{X/Y}\otimes_{\O_X}\L$ by letting $\nabla_r (\omega \sigma)=
d_r\omega\otimes\sigma - \omega \wedge \nabla_r (\sigma)$.

\proclaim {Definition} The curvature $K_r$ of a relative connection
$\nabla_r$ over $\L$ is the morphism of $\C$-modules
$$K_r=\nabla_r^2\colon \L\to \Omega^2_{X/Y} \otimes_{\O_X}\L\,.\tag 3.13$$
\endproclaim

By construction, if $K$ is the curvature of the ordinary connection
$\nabla$ over $\L$ which induces $\nabla_r$, one has $K_r=(p\otimes
\operatorname{Id})\circ K$.
In particular, $K_r$ is an $\O_X$-linear morphism,
and therefore it determines
a global section $K_r$ of $ \Omega_{X/Y}^2$. By Eq\. (2.1)
$K_r$ is also closed under
the relative differential, and then $K_r\in\Gamma
(Y,\pi_\ast\Zc_{X/Y}^2)$.
The projection $[K_r]=[p(K)]=p([K])\in\Gamma (Y,DR^2_{X/Y})$
does not depend on the relative connection over $\L$ (because $[K]$ does
not),
and from (3.10) one concludes that
$$c_1[\L]= \frac{i}{2\pi} [K_r] \in \Gamma (Y,DR^2_{X/Y})\,.\tag 3.14$$

If $\pi\colon (X,\O_X)\to (Y,\O_Y)$ is a family of Riemann surfaces,
Proposition 2.5 yields the identification $[K_r] = \intXY K_r
\in \Gamma (Y,\Cc_Y)$, which together with (3.14) proves a relative
Gauss-Bonnet theorem.
\proclaim {Theorem} Let $\pi\colon (X,\O_X)\to (Y,\O_Y)$
be a family of Riemann surfaces, $\L$ a line bundle over $(X,\O_X)$,
let $[\L]$ be its image in $\picXY$, and $K_r$ the curvature of any
relative connection over $\L$; then
$$\hbox to\hsize{$\phantom{\hbox{\cub}}$\hfill${\dsize
c_1([\L])= \frac{i}{2\pi} \intXY K_r \,.}$\hfill\cub}$$
\endproclaim
\proclaim{Corollary}  Assume that $\phi:\pic X \to \picXY$ is
surjective, and
let $\lambda\in\picXY$. Then $c_1(\lambda)= \frac{i}{2\pi} \intXY K_r$,
where
$K_r$ is the curvature of a relative connection on any line bundle on
$(X,\O_X)$
whose image in $\picXY$ is $\lambda$.\qed\endproclaim

\obligedskip\subheading{Acknowledgements} We would like to express warm
thanks
to C\. Bartocci and D\. Hern\'andez Ruip\'erez for their contribution
to the initial stage of this work. We also thank
D\. Hern\'andez Ruip\'erez and J\. Mu\~noz Porras for enlightening
comments on relative duality.

\obligedskip
\enddocument
\bye